\documentclass[conference]{IEEEtran}
\IEEEoverridecommandlockouts

\usepackage{cite}
\usepackage{amsmath,amssymb,amsfonts}
\usepackage{algorithmic}
\usepackage{graphicx}
\usepackage{textcomp}
\usepackage[table]{xcolor}
\usepackage{subcaption}
\usepackage{multirow}
\usepackage{booktabs}
\usepackage{tikz}
\usepackage{tabularx}
\usepackage{arydshln}
\usepackage{threeparttable}
\definecolor{LightCyan}{rgb}{0.88,1,1}
\definecolor{LightGreen}{rgb}{0.9,1,0.9}
\usepackage{hyperref}

\def\BibTeX{{\rm B\kern-.05em{\sc i\kern-.025em b}\kern-.08em
    T\kern-.1667em\lower.7ex\hbox{E}\kern-.125emX}}
\begin{document}

\title{DiffRhythm+: Controllable and Flexible Full-Length Song Generation with Preference Optimization}

\author{
 \IEEEauthorblockN{
  Huakang Chen\IEEEauthorrefmark{2}, 
  Yuepeng Jiang\IEEEauthorrefmark{2}, 
  Guobin Ma\IEEEauthorrefmark{2}, 
  Chunbo Hao\IEEEauthorrefmark{2}\\
  Shuai Wang\IEEEauthorrefmark{3},
  Jixun Yao\IEEEauthorrefmark{2}, 
  Ziqian Ning\IEEEauthorrefmark{2}, 
  Meng Meng\IEEEauthorrefmark{4}, 
  Jian Luan\IEEEauthorrefmark{4}, 
  Lei Xie\textsuperscript{*}\IEEEauthorrefmark{2}}\thanks{* Corresponding author.}
 \IEEEauthorblockA{\IEEEauthorrefmark{2}Audio, Speech and Language Processing Lab (ASLP@NPU)}
 \IEEEauthorblockA{\IEEEauthorrefmark{3}School of Intelligence Science and Technology, Nanjing University, Suzhou, China}
 \IEEEauthorblockA{\IEEEauthorrefmark{4}MiLM Plus, Xiaomi Inc.} 
 \IEEEauthorblockA{huakang@mail.nwpu.edu.cn, lxie@nwpu.edu.cn}
}

\maketitle

\begin{abstract}
Songs, as a central form of musical art, exemplify the richness of human intelligence and creativity. While recent advances in generative modeling have enabled notable progress in long-form song generation, current systems for full-length song synthesis still face major challenges, including data imbalance, insufficient controllability, and inconsistent musical quality. DiffRhythm, a pioneering diffusion-based model, advanced the field by generating full-length songs with expressive vocals and accompaniment. However, its performance was constrained by an unbalanced model training dataset and limited controllability over musical style, resulting in noticeable quality disparities and restricted creative flexibility.
To address these limitations, we propose DiffRhythm+, an enhanced diffusion-based framework for controllable and flexible full-length song generation. DiffRhythm+ leverages a substantially expanded and balanced training dataset to mitigate issues such as repetition and omission of lyrics, while also fostering the emergence of richer musical skills and expressiveness. The framework introduces a multi-modal style conditioning strategy, enabling users to precisely specify musical styles through both descriptive text and reference audio, thereby significantly enhancing creative control and diversity. We further introduce direct performance optimization aligned with user preferences, guiding the model toward consistently preferred outputs across evaluation metrics.
Extensive experiments demonstrate that DiffRhythm+ achieves significant improvements in naturalness, arrangement complexity, and listener satisfaction over previous systems. Audio samples and code are available at \url{https://longwaytog0.github.io/DiffRhythmPlus/} and \url{https://github.com/ASLP-lab/DiffRhythm}.
\end{abstract}


\section{Introduction}
\label{sec:intro}

Music plays an essential role in human culture, serving as a profound medium for emotional expression, cultural transmission, and social interaction throughout history \cite{music_history1,music_evolution}. Despite its ubiquity, traditional music composition remains technically demanding, typically requiring extensive expertise and substantial resources, thus limiting accessibility for broader audiences \cite{owsinski2016music}. Recent advancements in neural generative models have significantly lowered barriers, enabling more accessible, efficient, and interactive music creation. These developments facilitate various applications, such as personalized music generation for short videos, film scoring, educational tools, and therapeutic practices \cite{novice,music_gen_review}.

Research within the domain of music generation generally encompasses three primary areas: Singing Voice Synthesis (SVS), text-to-music generation, and lyric-to-song generation. SVS generates expressive, human-like singing voices from provided lyrics and musical scores \cite{Liu2021DiffSingerSV,hifisinger,visinger,hiddensinger,toksing}, supporting applications such as virtual singers, artist voice cloning, and assistive tools. In contrast, text-to-music generation creates instrumental music conditioned on textual descriptions, mood cues, or prompts \cite{melody,musicldm,musiclm,melodist,singsong,riffusion}. However, both approaches, when used independently, present inherent limitations. SVS models typically produce only vocal tracks without instrumental accompaniment, whereas text-to-music systems generate instrumental tracks lacking vocal melodies and lyrics. In real-world compositions, vocals and accompaniment intertwine to create rich semantic and acoustic coherence, posing substantial challenges for comprehensive song generation.

Song generation specifically addresses this challenge by synthesizing complete songs, including both vocals and instrumentals, directly from raw lyrics and style prompts. Current methods generally adopt either autoregressive language model (LM)-based approaches or diffusion-based techniques. LM-based approaches \cite{JukeBox,melodylm,songcreator,songgen,YuE} formulate song generation as sequential modeling tasks, employing large-scale language models to produce musical tokens from lyrics in an autoregressive manner. While effective at capturing complex musical structures and multi-modal conditioning, these models suffer from slow inference, hindering real-time and interactive applications.

In contrast, diffusion-based approaches \cite{musicldm,noise2music,stable_audio,mustango} generate song by iteratively denoising latent representations, providing superior controllability, high audio fidelity, and robust long-term coherence. Diffusion-based approaches excel at flexible conditioning on various musical attributes, enabling nuanced control over melody, rhythm, and timbre. Furthermore, diffusion-based methods naturally support detailed editing, inpainting, and style transfer, making them highly versatile and expressive for song generation tasks. 
DiffRhythm \cite{DiffRhythm2025} is the first open-source diffusion-based model designed for full-length song generation, capable of synthesizing both vocal and accompaniment tracks for songs up to more than 4 minutes. 
DiffRhythm stands out from previous methods that rely on complex pipelines or can only generate short or isolated musical elements, by enabling fast, end-to-end generation of complete songs. With only lyrics and a style prompt as input, it can deliver a full-length song in as little as ten seconds, making it ideal for real-time and interactive music applications.

Despite its strengths, DiffRhythm exhibits several notable limitations. First, the training dataset used in DiffRhythm is imbalanced, with an approximate ratio of 3:6:1 among Chinese songs, English songs, and instrumental music. This imbalance results in significantly poorer performance for Chinese songs compared to English ones, often causing repetitive or missing lyrics. Second, recent studies such as YuE~\cite{YuE} have shown that large-scale training data is crucial for improving musical complexity and expressive quality. Consequently, the relatively limited scale of DiffRhythm’s dataset constrains its ability to produce outputs with rich structural complexity, detailed arrangements, and nuanced instrumental expression. Moreover, DiffRhythm’s style control is restricted to audio-based prompts, limiting the flexibility and precision of stylistic variations. Finally, the model occasionally demonstrates instability, generating outputs with inconsistent clarity and coherence.

To address these shortcomings, we propose DiffRhythm+, an enhanced song generation framework that significantly improves style control flexibility and aligns generated outputs closely with listener preferences. DiffRhythm+ substantially elevates the quality, diversity, and controllability of generated music. The key contributions of DiffRhythm+ are summarized as follows:

\begin{itemize} 
    \item We propose DiffRhythm+, an advanced framework for full-length song generation that integrates multi-modal style control, large-scale balanced training data, and listener preference optimization, significantly enhancing the quality and creative versatility of AI-generated music. 
    \item We incorporate preference optimization guided by music aesthetic evaluation models, aligning generated outputs more closely with human preferences and enhancing musical expressiveness, arrangement complexity, and listener satisfaction. 
    \item We employ MuLan \cite{mulan,muq}, a multi-modal style extractor that enables fine-grained style conditioning through both textual descriptions and audio prompts, greatly increasing flexibility and creative control. 
\end{itemize}

\begin{figure*}[!htbp]
\centering
\includegraphics[scale=0.5]{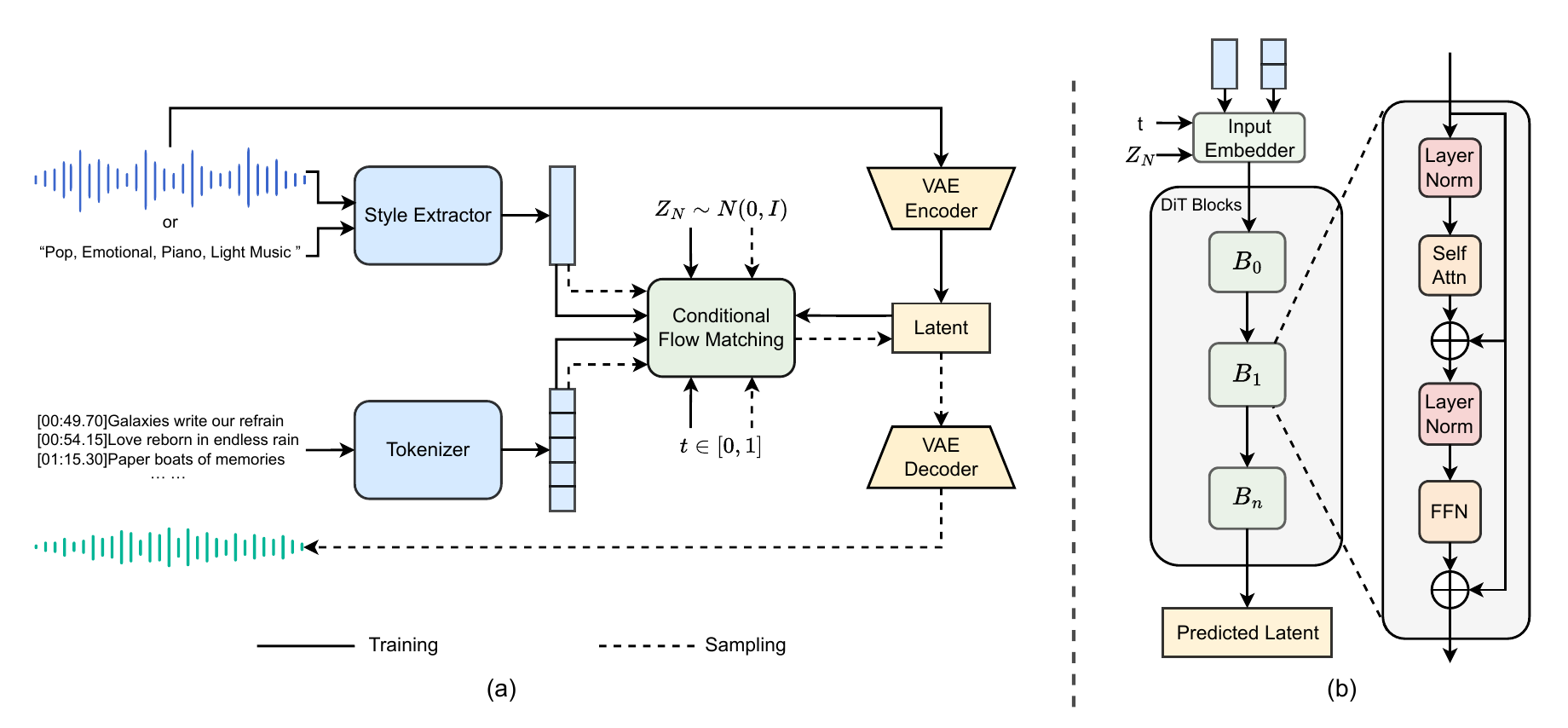}
\caption{(a) Overview of the DiffRhythm+ architecture. Style descriptions are encoded by a style extractor, and lyrics are tokenized by a text tokenizer; both resulting embeddings serve as external control signals to the DiT-based conditional flow matching module. A VAE module decodes the resulting latent variables to produce the final audio output.
(b) Detailed structure of the conditional flow matching module based on DiT.}
\vspace{-10pt}
\label{fig:model}
\end{figure*}

\section{Related Work}
\label{sec:related work}
\subsection{Language Model-Based Song Generation}
Early work in language model-based song generation, such as Jukebox \cite{JukeBox}, leveraged hierarchical VQ-VAE architectures combined with transformer decoder architecture to generate complete songs from textual inputs. 
More recent models, such as MelodyLM \cite{melodylm}, SongCreator \cite{songcreator}, and SongGen \cite{songgen}, introduce multi-stage tokenization schemes and incorporate multi-modal conditioning to enhance the alignment between generated audio components and input lyrics. These approaches represent meaningful progress in modeling musical structure, lyric expressiveness, and stylistic control. However, they still face significant challenges in generating long-form, high-fidelity audio with consistent coherence across sections.
A notable advancement in this line of work is YuE \cite{YuE}, which achieves impressive performance in long-form lyric-to-song generation by scaling up both model capacity and training data. Nevertheless, its reliance on large-scale autoregressive modeling leads to slow inference and high computational demands, making it less suitable for interactive or resource-constrained deployment scenarios.

\subsection{Diffusion-Based Song Generation}

Diffusion-based generative models have shown potential in expressive and coherent long-form music generation. Recent works such as Noise2Music \cite{noise2music}, Stable Audio \cite{stable_audio}, MusicLDM \cite{musicldm}, and MUSTANGO \cite{mustango} adopt latent diffusion techniques operating on compressed audio representations. This design enables efficient synthesis while preserving high audio quality across extended durations.
Diffusion models also naturally support multi-modal conditioning and editing operations such as inpainting and continuation, making them especially suitable for creative tasks. For example, Music-ControlNet \cite{music_controlnet}, MusicGen \cite{musicgen}, and JASCO \cite{jasco} enable fine-grained control over musical attributes such as genre, instrumentation, and melody. In addition to instrumental generation, diffusion-based methods have also been applied to SVS and voice conversion \cite{ldmsvc, rthfsvc}.
Furthermore, several hybrid approaches have emerged that integrate diffusion and language modeling techniques. Systems like MeLoDy \cite{melody} and SongCreator \cite{songcreator} use diffusion modules for audio decoding while relying on autoregressive language models for symbolic generation. Although hybrid systems improve flexibility and overall quality, they often involve complex pipelines and inherit the slow inference speed of autoregressive components.

\section{Diffrhythm+}
\label{sec:proposed method}

\subsection{Overview}
\label{ssec:overview}

DiffRhythm+ extends the capabilities of DiffRhythm by enhancing controllability, expressiveness, and alignment quality in full-length song generation. The model comprises two main components: (1) a VAE module that encodes waveforms into latent representations and reconstructs them back to audio, and (2) a flow matching module based on a diffusion transformer (DiT), which models the distribution of musical latents in a non-autoregressive manner.

A key improvement in DiffRhythm+ lies in its style conditioning mechanism. Instead of directly encoding a fixed audio segment as in DiffRhythm, DiffRhythm+ employs MuLan to extract semantically rich and expressive style embeddings from audio prompts. This allows for more precise and flexible control over musical attributes such as genre and emotion.
DiffRhythm+ also refines the lyrics-to-latent alignment strategy by introducing multi-frame temporal perturbation \cite{DiffRhythm2025}. This technique improves robustness to noisy or imprecise timestamps by relaxing rigid alignments, enabling the model to generate more fluid and expressive musical phrasing even with imperfect lyric annotations.
In addition to these architectural improvements, DiffRhythm+ introduces a preference optimization stage to align generation more closely with human musical preferences—an element absent from the original DiffRhythm design.

\subsection{Detailed Architecture}
\label{ssec:detailed architecture}

As illustrated in Figure~\ref{fig:model}, the DiT is conditioned on three main features: a style prompt for controlling musical style, a timestep indicating the current diffusion step, and lyrics for vocal content control. Moreover, instead of the LSTM network used in DiffRhythm, DiffRhythm+ adopts MuLan as the style extractor. MuLan is a cross-modal representation model that aligns audio and text into a shared embedding space, enabling the model to accept both text and audio style prompts in a unified manner. Specifically, MuQ-MuLan \cite{muq} exhibits strong capability in extracting representative and expressive features from music, making it a valuable component in our system. This design not only facilitates more flexible and accurate style conditioning but also allows for direct multimodal control of song style during generation. For lyric encoding, we employ a G2P-based text tokenizer to convert lyrics into phoneme sequences, providing a linguistically informed representation for the model.

The style embedding extracted by MuLan, together with the timestep embedding, lyrics embedding, and noisy latent representations, are concatenated along the feature dimension and then fed into DiT blocks as conditioning inputs. This design allows DiffRhythm+ to flexibly incorporate multimodal style control and vocal content, enabling more expressive and controllable song generation.

\subsection{Training Objective}
\label{ssec:training obj}
DiffRhythm+ adopts conditional flow matching \cite{cfm} to model the generative process in the latent space. Specifically, the model learns a time-dependent vector field $v_\theta(t, y, c)$, parameterized by a neural network, which transforms samples from the prior distribution $p_0$ to the data distribution $p_1$, conditioned on multi-modal inputs $c$ (including lyrics and style features).

Formally, given a pair of samples $y^- \sim p_0$, $y^+ \sim p_1$, the conditional flow matching objective is:
\begin{align}
\label{eq:cfm loss}
\mathcal{L}_{\mathrm{CFM}}(\theta) = \mathbf{E}_{t \sim [0, 1], z \sim q(z), y \sim p_{t}(y|z)} \left[ \left| v_\theta(t, y, c) - u_t(y|z) \right|^2 \right],
\end{align}
where $q(z)$ is the coupling distribution of $(y^-, y^+)$, $p_t(y|z)$ is the conditional path between $y^-$ and $y^+$, and $u_t(y|z) = y^+ - y^-$ denotes the straight path velocity.
In DiffRhythm+, the coupling $q(z)$ is implemented as independent sampling of $y^- \sim p_0$ and $y^+ \sim p_1$, and $c$ encodes the lyric and style prompt features. This framework enables the model to robustly learn the transformation from noise to song latent representations, facilitating high-quality, conditional song generation.

\begin{figure*}[!htbp]
  \centering
  \begin{subfigure}[b]{0.32\linewidth}
    \includegraphics[width=\linewidth]{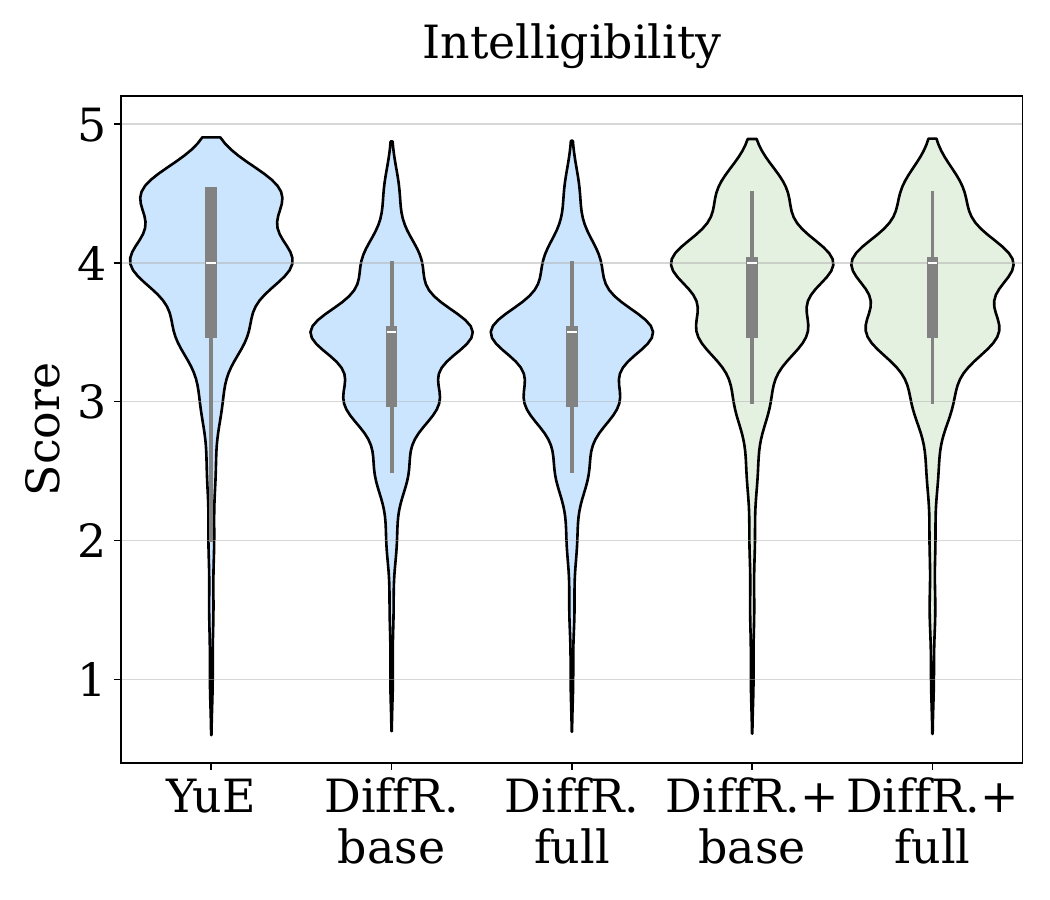}
  \end{subfigure}
  \begin{subfigure}[b]{0.32\linewidth}
    \includegraphics[width=\linewidth]{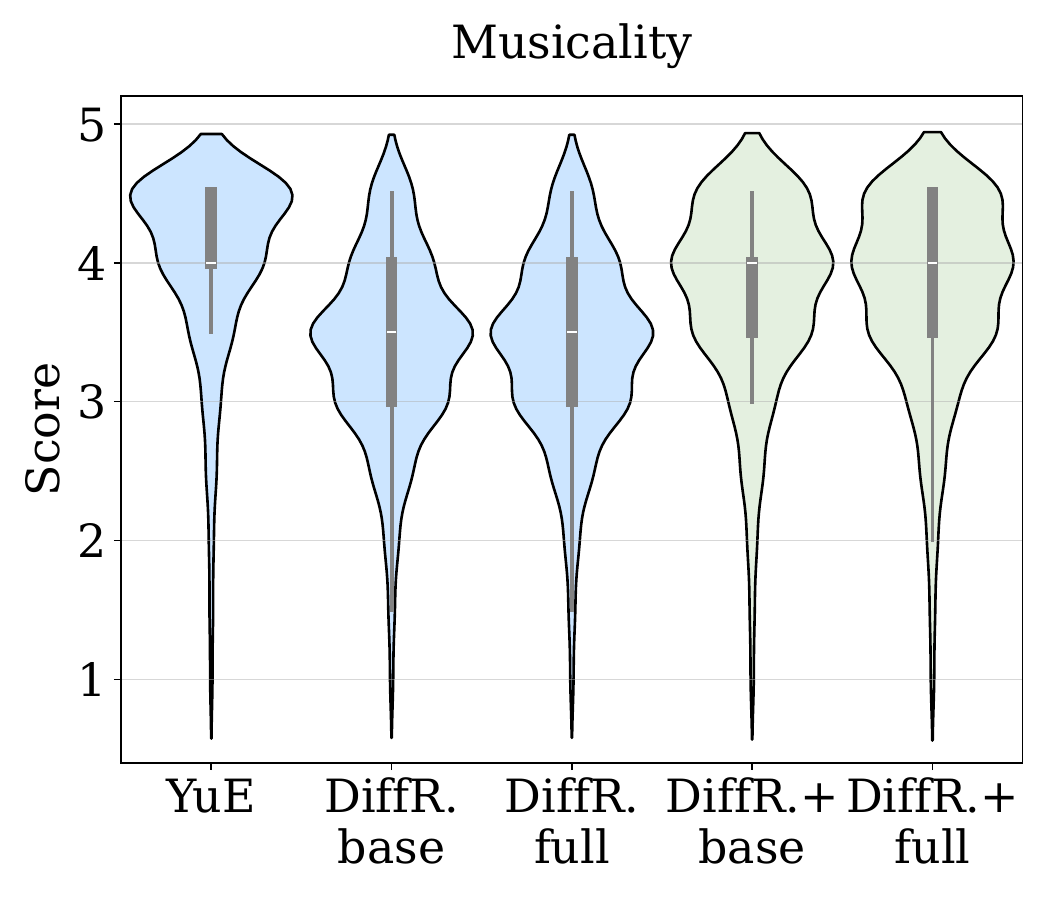}
  \end{subfigure}
  \begin{subfigure}[b]{0.32\linewidth}
    \includegraphics[width=\linewidth]{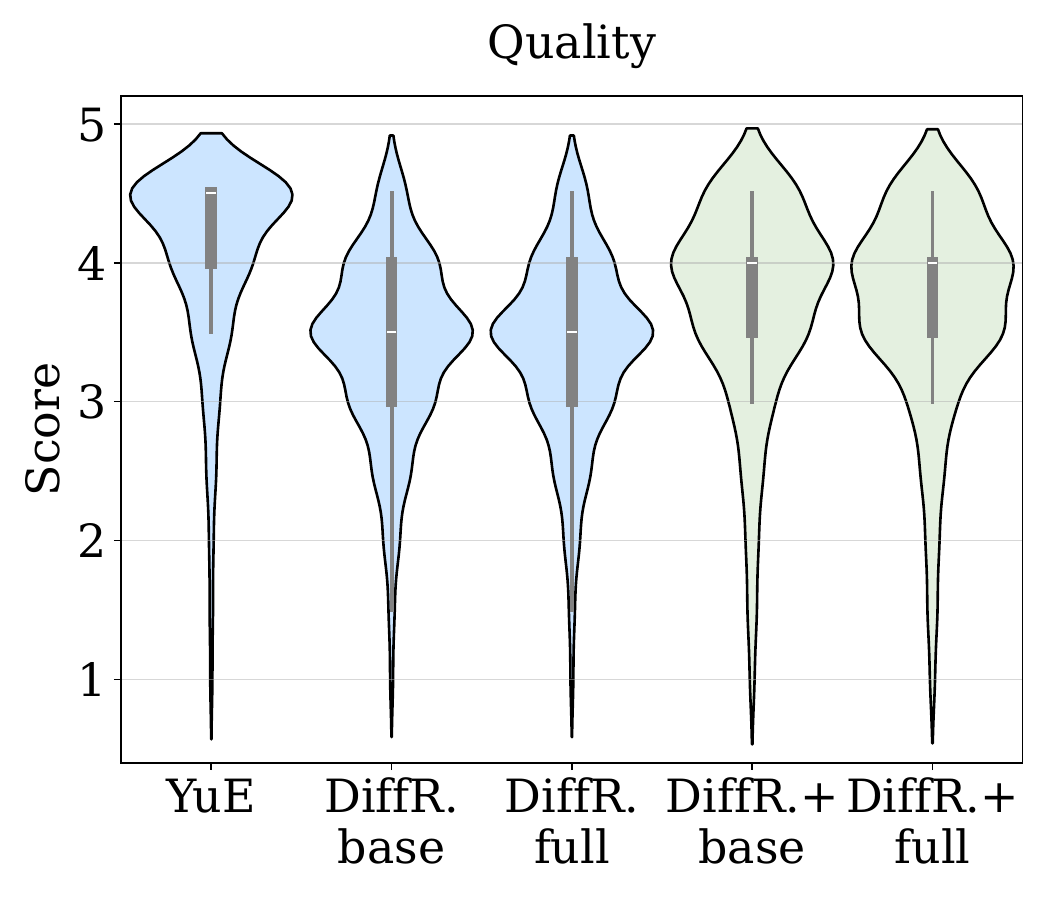}
  \end{subfigure}
  \caption{Violin plots of MOS results for intelligibility, musicality, and quality evaluation.}
  \label{fig:subjective:mos}
\vspace{-10pt}
\end{figure*}

\subsection{Preference Optimization}
\label{ssec:preference opt}
\textbf{Theoretical Motivation}\quad
Inspired by Tango2’s preference optimization in diffusion-based models \cite{tango2}, DiffRhythm+ integrates Direct Preference Optimization (DPO) \cite{dpo} into the proposed generation framework, enabling explicit alignment with human preferences. 
The conventional DPO method proposed for autoregressive language models seeks to optimize the following objective:
\begin{align}
\label{eq:rlhf}
    \max_{\pi_{\theta}} \mathbb{E}_{\tau \sim \mathcal{D}, x \sim \pi_{\theta}(x|\tau)} \left[ r_{\phi}(\tau, x) \right] - \beta D_{KL} \left[ \pi_{\theta}(x|\tau) \parallel \pi_{\text{ref}}(x|\tau) \right].
\end{align}
Here, $\tau$ denotes the input prompt, $x$ the generated output, $r_{\phi}$ the reward model (often derived from preference metrics), and $\pi_{\text{ref}}$ a reference model. In practice, maximum likelihood estimation can be used with a preference dataset to obtain the negative log likelihood (NLL) loss:
\begin{align}
    \mathcal{L}_R(r_{\phi}, \mathcal{D}) = -\mathbb{E}_{(\tau, x^w, x^l) \sim \mathcal{D}} \left[ \log \sigma (r_{\phi}(\tau, x^w) - r_{\phi}(\tau, x^l)) \right].  \label{eq:NLL}
\end{align}

However, for diffusion models, the optimization target extends to the entire diffusion trajectory. The objective is defined as:
\begin{flalign}
    \max_{\pi_\theta} \mathbb{E}_{\tau \sim \mathcal{D}, x_{0:N}\sim \pi_\theta(x_{0:N} | \tau)} &[r(\tau, x_0)] \nonumber \\
    - \beta D_\text{KL} &[ \pi_\theta(x_{0:N} | \tau) || \pi_\text{ref}(x_{0:N}|\tau)], \label{eq:diff-obj}
\end{flalign}
with the reward $r(\tau, x_0)$ defined as an expectation over the full diffusion path:
\begin{align}
    r(\tau, x_0) := \mathbb{E}_{\pi_\theta(x_{1:N} | x_0, \tau)} &[R(\tau, x_{0:N})]. \label{eq:diff-reward}
\end{align}

Through Jensen’s inequality and standard approximations in latent diffusion models (LDM), the final form of the diffusion-DPO loss is given as an L2 noise-estimation loss:
\begin{flalign}
    \mathcal{L}_\text{DPO-Diff} := &- \mathbb{E}_{n, \epsilon^w, \epsilon^l} \log \sigma(-\beta N \omega(\lambda_n) (||\epsilon_n^w - \hat\epsilon_\theta^{(n)}(x_n^w, \tau)||_2^2 \nonumber \\ &- ||\epsilon_n^w - \hat\epsilon_\text{ref}^{(n)}(x_n^w, \tau)||_2^2 \nonumber \\ &- (||\epsilon_n^l - \hat\epsilon_\theta^{(n)}(x_n^l, \tau)||_2^2 - ||\epsilon_n^l - \hat\epsilon_\text{ref}^{(n)}(x_n^l, \tau)||_2^2)), \label{eq:DPO-Diff}
\end{flalign}
where $(\tau, x_0^w, x_0^l)$ denotes the input prompt and the preferred/less-preferred samples (winner and loser).

\textbf{Win-Lose Pair Construction in DiffRhythm+}\quad
We employ specialized aesthetic metrics to conduct win-lose $(x_0^w, x_0^l)$ pairs for preference optimization in DiffRhythm+. Specifically, we employ both SongEval \cite{songeval} and Audiobox-aesthetic \cite{audiobox} as automated music aesthetic evaluation models to score generated songs.
Regarding metric models, we note that Audiobox is designed for general audio aesthetic evaluation, but recent work (e.g., YuE) has shown that its scores may not always align with human ratings. Therefore, for song-level evaluation, we primarily rely on the recently open-sourced SongEval tool, which is specifically designed for aesthetic evaluation of songs. However, as SongEval cannot score instrumental-only music, we use Audiobox for such cases. To maintain scoring consistency, Audiobox scores (originally in the 1–10 range) are linearly mapped to SongEval’s 1–5 scale.
For each batch of model-generated samples, the best and worst outputs are first identified based on their scores, providing candidate win-lose pairs. To ensure meaningful learning, we adopt a threshold-based filtering strategy: only pairs with a score gap exceeding a preset threshold are retained as DPO training data. 

\begin{table}[ht]
\centering
\renewcommand\arraystretch{1.3}
\caption{Comparison of various music generation models across multiple metrics. DiffR. denotes DiffRhythm and DiffR.$^+$ denotes DiffRhythm+; this abbreviation is used throughout all subsequent tables and figures for brevity.}
\label{tab:object:model}
\begin{tabular}{lccccc}
\hline
\multirow{2}{*}{Metric} 
  & \multicolumn{2}{c}{\textbf{Distri. Match}} 
  & \textbf{Align.} 
  & \textbf{Intelli.}
  & \textbf{Comp. Eff.} \\ \cline{2-3}
  & KL$\downarrow$ & FAD$\downarrow$ 
  & CLaMP 3$\uparrow$ & PER$\downarrow$ & RTF$\downarrow$ \\
\hline
SongLM            & 0.736  & 1.921  & 0.101  & 21.35\% & 1.717 \\
YuE               & \textbf{0.372}  & \textbf{1.624}  & \textbf{0.240}  & 15.14\% & 10.385 \\
\hline
\makebox[1.5cm][l]{DiffR.$^+$\hfill base}  & \underline{0.488}  & \underline{1.835}  & 0.152  & \textbf{14.85\%} & \underline{0.036} \\
\makebox[1.5cm][l]{DiffR.\hfill base}   & 0.723  & 2.113  & 0.103  & 17.47\% & \textbf{0.034} \\
\makebox[1.5cm][l]{DiffR.$^+$\hfill full}  & 0.512  & 1.872  & \underline{0.165}  & \underline{14.96\%} & 0.039 \\
\makebox[1.5cm][l]{DiffR.\hfill full}   & 0.741  & 2.251  & 0.112  & 18.02\% & 0.037 \\
\hline
\end{tabular}
\end{table}

\section{Experimental Setup}
\label{sec:exp setup}

\subsection{Dataset}
\label{ssec:dataset}

\begin{table*}[ht]
\centering
\renewcommand\arraystretch{1.3}
\caption{Quantitative comparison of music generation models using aesthetic evaluation criteria. Best results are highlighted in \textbf{bold}, and second-best results are \underline{underlined}.}
\label{tab:object:aesthetic}
\begin{tabularx}{\textwidth}{lXXXX@{\hskip 15pt}c@{\hskip 15pt}XXXXX}
\hline
\multirow{2}{*}{Metric Model} 
  & \multicolumn{4}{c}{\textbf{AudioBox-aesthetic}} 
  & & \multicolumn{5}{c}{\textbf{SongEval}} \\ \cline{2-5} \cline{7-11}
  & CE$\uparrow$ & CU$\uparrow$ & PC$\uparrow$ & PQ$\uparrow$
  & & Coh$\uparrow$ & Mem$\uparrow$ & NVBP$\uparrow$ & CSS$\uparrow$ & OM$\uparrow$ \\ \hline
GT     & 7.159 & 7.578 & 5.975 & 7.787 & & 4.078 & 4.048 & 3.786 & 3.780 & 3.823 \\
\hline
SongLM & 6.538 & 7.232 & 5.614 & 7.412 & & 2.588 & 2.474 & 2.278 & 2.233 & 2.412 \\
YuE    & 7.115 & \underline{7.543} & \underline{6.280} & \underline{7.894} & & 3.722 & \textbf{3.714} & \textbf{3.288} & \underline{3.418} & \textbf{3.458} \\
\hline
\makebox[1.5cm][l]{DiffR.$^+$\hfill base} & \underline{7.345} & \textbf{7.769} & \textbf{6.283} & \textbf{7.978} & & \textbf{3.768} & \underline{3.682} & \underline{3.203} & 3.414 & 3.287 \\
\makebox[1.5cm][l]{DiffR.\hfill base}  & 6.224 & 7.163 & 5.351 & 7.421 & & 2.634 & 2.417 & 2.215 & 2.261 & 2.281 \\
\makebox[1.5cm][l]{DiffR.$^+$\hfill full} & \textbf{7.443} & 7.518 & 6.224 & 7.852 & & \underline{3.738} & 3.675 & 3.153 & \textbf{3.421} & \underline{3.395} \\
\makebox[1.5cm][l]{DiffR.\hfill full}  & 6.481 & 6.912 & 5.441 & 7.441 & & 2.612 & 2.531 & 2.334 & 2.185 & 2.331 \\
\hline
\end{tabularx}
\vspace{-10pt}
\end{table*}

For model pre-training, we first collect 300,000 hours of raw music recordings from the internet and apply a rule-based filtering pipeline inspired by DiffRhythm. The resulting dataset contained $\sim$120,000 hours of high-quality music, balanced across Chinese songs, English songs, and instrumental tracks in a 2:2:1 ratio. This curated corpus was used to pre-train DiffRhythm+.
To further enhance generation quality, we apply audio quality filtering using Audiobox-aesthetic and SongEval. A small high-quality subset is manually selected to determine score-based thresholds: Audiobox alone for instrumentals, and a weighted combination of Audiobox and SongEval for vocal tracks. After filtering, 25,000 hours of top-quality music remained, which were used for supervised fine-tuning (SFT).

\subsection{Baseline Systems}
\label{ssec:baseline systems}

To evaluate the performance of DiffRhythm+, we conduct experiments with several baseline systems:

\textbf{SongLM} \cite{songeditor} employs a two-stage approach that integrates an autoregressive LM to generate discrete semantic tokens conditioned on lyrics and short acoustic prompts and a diffusion-based generator to reconstruct music, respectively.  

\textbf{YuE} \cite{YuE} comprises a two-stage, dual language model framework tailored for lyrics-to-song generation. Stage one involves a track-decoupled next-token prediction, separately modeling vocals and accompaniment tokens. Stage two involves residual modeling for precise audio reconstruction.

\textbf{DiffRhythm} \cite{DiffRhythm2025} introduces a latent diffusion-based model focusing on fast, non-autoregressive generation of full-length songs. It employs a DiT module conditioned on lyrical content and style prompts, along with a sentence-level lyrics alignment mechanism, improving vocal intelligibility and musical coherence across the generated track.

\subsection{Evaluation Metrics}
\label{ssec:eval metrics}
\textbf{Objective Evaluation}\quad
For objective evaluation, we retain the primary metrics used in DiffRhythm, including Fréchet Audio Distance (FAD) \cite{fad}, Phoneme Error Rate (PER), and Real-Time Factor (RTF). Additionally, we incorporate Kullback-Leibler (KL) divergence, inspired by YuE, and pair it with FAD as distribution matching metrics to assess how well the generated audio matches the real data distribution. For alignment evaluation, we use CLaMP 3 \cite{clamp3}, which measures semantic alignment between style conditions and generated audio. To comprehensively evaluate content quality, we adopt Audiobox-aesthetic~\cite{audiobox} and SongEval~\cite{songeval} as content-based metrics. Audiobox-aesthetic covers production quality (PQ; clarity, fidelity, dynamics, and spatialization), production complexity (PC; richness of audio components), content enjoyment (CE; subjective appeal and artistic value), and content usefulness (CU; potential for reuse in content creation). SongEval further assesses overall coherence (Coh), memorability (Mem), the naturalness of vocal breathing and phrasing (NVBP), clarity of song structure (CSS), and overall musicality (OM), thus reflecting both technical and aesthetic aspects of the generated music. All evaluations follow established protocols, and RTF is measured on an Nvidia RTX 4090 to demonstrate computational efficiency.

\textbf{Subjective Evaluation}\quad
We conduct mean opinion score (MOS) listening tests for subjective evaluation. A total of 30 listeners participated in the assessment, including 15 music experts and 15 participants without a formal musical background. Each participant rated the generated song samples on a scale from 1 to 5 across three aspects: musicality, audio quality, and intelligibility.

\subsection{Model Configuration}
DiffRhythm+ reuses the VAE architecture and pre-trained weights from DiffRhythm. The overall model has approximately 1.1B parameters. The DiT backbone comprises 16 LLaMA decoder layers \cite{llama} with a 2048-dimensional hidden size and 32-head self-attention. Training is performed in two stages: initial pre-training on shorter sequences ($L_{max}=2048$) followed by fine-tuning on longer sequences ($L_{max}=6144$). The diffusion process utilizes an Euler ODE solver with 32 steps and classifier-free guidance (CFG) \cite{cfg} with a scale of 4 during inference. Independent 20\% dropout is applied to both lyrics and style prompts.
For DPO training, the $\beta$ parameter in the DPO loss (Eq.~(\ref{eq:DPO-Diff})) is set to 2000. Win-lose pairs are constructed using a score gap threshold of 0.4, and the winner’s score must exceed 3 on the normalized five-point SongEval scale. The DPO stage uses a per-GPU batch size of 8 across eight GPUs.

We adopt the AdamW optimizer ($\beta_1=0.9$, $\beta_2=0.95$) throughout. The learning rate is set to $1 \times 10^{-4}$ for pre-training, $1 \times 10^{-5}$ for SFT, and $1 \times 10^{-6}$ for DPO. SFT is conducted for 10 epochs on the filtered SFT dataset, while DPO is performed for 8 epochs on the preference data. An exponential moving average (EMA) of model weights is maintained with a decay rate of 0.99, updated every 100 batches.

\begin{table*}[ht]
\centering
\caption{Evaluation score distribution (number of samples in each score range) across all ablation settings. For each block, the best results (based on the highest \textbf{mean} score) are highlighted in \colorbox{LightGreen}{green}. Sample counts separated by / denote results for Chinese/English, respectively.}
\begin{tabular}{llcccccc}
\toprule
Ablation Category & Setting & [1--2]$\downarrow$ & [2--3]$\downarrow$ & [3--4]$\uparrow$ & [4--5]$\uparrow$ & Mean$\uparrow$ & {[}3--5{]} \%$\uparrow$ \\
\midrule
\multirow{2}{*}{\textbf{Data Scale \& Balance}} 
 & 60,000 hours (ZH:EN = 1:1)   & {33 / 32} & {32 / 28} & {35 / 40} & 0 & 2.31 & 37.5\% \\
 & 120,000 hours (ZH:EN = 1:1)  & \cellcolor{LightGreen}{2 / 1} & \cellcolor{LightGreen}58 / 51 & \cellcolor{LightGreen}{40 / 48} & \cellcolor{LightGreen}0 & \textbf{\cellcolor{LightGreen}2.80} & \cellcolor{LightGreen}44.0\% \\
 & 120,000 hours (ZH:EN = 1:2)  & 4 / 1 & 58 / 40 & 38 / 59 & 0 & 2.75 & 48.5\% \\
\midrule
\multirow{5}{*}{\textbf{Style Conditioning}} 
 & Audio Latent + LSTM & 14 & 117 & 69 & 0 & 2.63 & 34.5\% \\
 & Audio MuLan (Audio Prompt) & 4 & 87 & 107 & 2 & 2.85 & 54.5\% \\
 & Mixed MuLan (Audio Prompt) & \cellcolor{LightGreen}4 & \cellcolor{LightGreen}88 & \cellcolor{LightGreen}107 & \cellcolor{LightGreen}1 & \textbf{\cellcolor{LightGreen}2.88} & \cellcolor{LightGreen}54.0\% \\
\noalign{\vskip 1pt}
\cdashline{2-8}
\noalign{\vskip 1pt}
 & T5 Embedding + LSTM & 30 & 119 & 51 & 0 & 2.46 & 25.5\% \\
 & Audio MuLan (Text Prompt)  & 71 & 108 & 21 & 0 & 2.23 & 10.5\% \\
 & Mixed MuLan (Text Prompt)  & \cellcolor{LightGreen}11 & \cellcolor{LightGreen}106 & \cellcolor{LightGreen}83 & \cellcolor{LightGreen}0 & \textbf{\cellcolor{LightGreen}2.75} & \cellcolor{LightGreen}41.5\% \\
\midrule
\multirow{4}{*}{\textbf{DPO \& Training Stage}} 
 & DiffR.  & 68 & 91 & 40 & 1 & 2.25 & 20.5\% \\
 & DiffR.$^+$ Pretrain & 3 & 109 & 88 & 0 & 2.80 & 44.0\% \\
 & DiffR.$^+$ SFT      & 3 & 78 & 110 & 9 & 2.86 & 59.5\% \\
 & DiffR.$^+$ DPO      & \cellcolor{LightGreen}1 & \cellcolor{LightGreen}36 & \cellcolor{LightGreen}145 & \cellcolor{LightGreen}18 & \textbf{\cellcolor{LightGreen}3.19} & \cellcolor{LightGreen}81.5\% \\
\midrule
\multirow{2}{*}{\textbf{DPO Winner}} 
 & GT Winner        & 7 & 64 & 127 & 2 & 2.94 & 64.5\% \\
 & Generated Winner & \cellcolor{LightGreen}1 & \cellcolor{LightGreen}36 & \cellcolor{LightGreen}145 & \cellcolor{LightGreen}18 & \textbf{\cellcolor{LightGreen}3.19} & \cellcolor{LightGreen}81.5\% \\
\midrule
\multirow{3}{*}{\textbf{DPO Training Epochs}} 
 & 4 Epoch  & 4 & 47 & 139 & 10 & 3.11 & 74.5\% \\
 & 8 Epoch  & \cellcolor{LightGreen}1 & \cellcolor{LightGreen}36 & \cellcolor{LightGreen}145 & \cellcolor{LightGreen}18 & \textbf{\cellcolor{LightGreen}3.19} & \cellcolor{LightGreen}81.5\% \\
 & 12 Epoch & 2 & 35 & 138 & 25 & 3.16 & 81.5\% \\
\midrule
\multirow{3}{*}{\textbf{Win-Lose Score Gap}} 
 & Gap = 0   & 4 & 38 & 138 & 20 & 3.15 & 79.0\% \\
 & Gap = 0.4 & \cellcolor{LightGreen}1 & \cellcolor{LightGreen}36 & \cellcolor{LightGreen}145 & \cellcolor{LightGreen}18 & \textbf{\cellcolor{LightGreen}3.19} & \cellcolor{LightGreen}81.5\% \\
 & Gap = 0.8 & 4 & 62 & 121 & 13 & 3.04 & 67.0\% \\
\bottomrule
\end{tabular}
\label{tab:ablation}
\vspace{-10pt}
\end{table*}

\section{Experimental Results}
\label{sec:exp res}
\subsection{Subjective Evaluation}
\label{ssec:human eval}
To comprehensively assess the subjective performance of DiffRhythm+ compared to baseline systems, we conduct a subjective evaluation focusing on three core musical aspects: intelligibility, musicality, and quality. The outputs of each system were rated on a 1–5 scale, and the aggregated results are visualized with violin plots in Figure \ref{fig:subjective:mos}.
We can find that YuE achieves the highest MOS across all aspects, which can be attributed to YuE’s larger model size (3B parameters), its extensive training data (650,000 hours), and the use of explicit vocal-accompaniment separation during training. Nevertheless, DiffRhythm+ exhibits substantial improvements over its predecessor, DiffRhythm, and outperforms other baseline models in most aspects. Both the base and full-length variants of DiffRhythm+ achieve higher median and upper-quartile scores compared to earlier versions, particularly in intelligibility and perceived audio quality, highlighting the benefits of the expanded and balanced training set as well as the improved model architecture and preference optimization strategy.

\subsection{Objective Evaluation}
\label{ssec:auto eval}

As shown in Table~\ref{tab:object:model}, DiffRhythm+ demonstrates substantial improvements over the original DiffRhythm across all objective metrics. For distribution matching, DiffRhythm+ (base: 0.488/1.835; full: 0.512/1.872) achieves much better KL divergence and FAD scores than the original DiffRhythm, approaching YuE (0.372/1.624). In terms of alignment (CLaMP 3) and intelligibility (PER), DiffRhythm+ achieves the lowest PER (14.85\%, 14.96\%) among baseline systems, even slightly outperforming YuE (15.14\%). Importantly, DiffRhythm+ achieves superior efficiency (RTF: 0.036–0.039) compared to YuE (10.385), demonstrating strong practical potential for fast and resource-efficient song generation.

Table~\ref{tab:object:aesthetic} reports musical quality via SongEval and AudioBox-aesthetic. On SongEval, which captures coherence, memorability, phrasing, structure, and overall musicality, DiffRhythm+ achieves performance on par with YuE. In particular, DiffRhythm+ (base) attains the highest coherence (Coh) and is competitive or superior on most other dimensions, while DiffRhythm+ (full) also closely matches YuE across the board. AudioBox scores favor DiffRhythm+ among baseline models, though discrepancies with human evaluations suggest its limitations as a preference proxy. Interestingly, several models exceed ground-truth scores on some AudioBox metrics, highlighting its potential overestimation. Overall, SongEval aligns more closely with subjective quality, and results affirm that DiffRhythm+ delivers strong musicality, structure, and naturalness, narrowing the gap with leading systems like YuE.

\subsection{Ablation Studies}
\label{ssec:ablations}

We conduct ablation studies to assess the effectiveness of data scale, data balance, style conditioning strategies, and preference optimization design. Results are summarized in Table~\ref{tab:ablation}. We observe that expanding the training data significantly improves output quality and consistency, and maintaining a balanced ratio between Chinese and English further enhances multilingual performance, highlighting the importance of both data scale and balance. For style conditioning, replacing raw audio latent embeddings (as used in DiffRhythm) with MuLan-based style features brings notable gains. Here, for example, \textit{Audio MuLan (Text Prompt)} refers to models trained with audio MuLan embeddings but inferred with text prompts, while \textit{Mixed MuLan (Audio Prompt)} indicates training with both audio and text MuLan embeddings and inference with audio prompts. Training with both audio and text MuLan embeddings (Mixed MuLan) achieves the best results, supporting robust multimodal style control and much better generalization to text prompts at inference compared to audio-only models. In contrast, using T5 embeddings with LSTM for style conditioning yields weaker performance, emphasizing the advantages of MuLan’s cross-modal alignment. Furthermore, applying preference optimization via DPO after SFT leads to substantial quality gains, with the best improvements seen when using generated outputs as DPO winners. Moderate hyperparameters (e.g., 8 epochs, 0.4 win-lose threshold) are most effective, while overly aggressive settings risk instability or overfitting.

\section{Conclusion and Future Work}
\label{sec:conclusion}

We propose DiffRhythm+, an upgraded diffusion-based framework for controllable, full-length lyric-to-song generation. Through dataset expansion and balancing, multimodal style conditioning, and preference optimization via DPO, DiffRhythm+ achieves notable improvements in intelligibility, musicality, and generation efficiency over its predecessor, while substantially closing the performance gap with state-of-the-art open-source models.

In future work, we plan to scale up model capacity, strengthen preference alignment, and explore more flexible and fine-grained control mechanisms. We also aim to reduce reliance on imprecise timestamps and incorporate richer structural cues to further improve the coherence and expressiveness of generated music.

\bibliographystyle{IEEEbib}
\bibliography{refs}

\end{document}